\documentclass[12pt,preprint]{aastex}

\shorttitle{CO Emission in High-$z$ QSOs}
\shortauthors{Hainline et al.}

\newcommand{\smm}{SMM~J04135+10277}
\newcommand{\rxj}{RX~J0911.4+0551}
\newcommand{\lbqs}{LBQS~0018-0220}
\newcommand{\msun}{M_{\sun}}
\newcommand{\lsun}{L_{\sun}}
\newcommand{\kms}{\textrm{km~s}^{-1}}
\newcommand{\jykms}{\textrm{Jy~km~s}^{-1}}
\newcommand{\hh}{\textrm{H}_{2}}
\newcommand{\mpc}{\textrm{Mpc}}
\newcommand{\Kkmspc}{\textrm{K}\,\,\textrm{km~s}^{-1}\,\textrm{pc}^{2}}
\newcommand{\mjybeam}{\textrm{mJy~beam}^{-1}}
\newcommand{\jybeam}{\textrm{Jy~beam}^{-1}}

\begin{document}

\title{A Study of CO Emission in High Redshift QSOs
 Using the Owens Valley Millimeter Array}

\author{Laura J. Hainline and N. Z. Scoville}
\affil{Dept.\ of Astronomy, California Institute
 of Technology, Mail Code 105-24, Pasadena, CA 91125, USA}
\email{ljh@astro.caltech.edu, nzs@astro.caltech.edu}

\author{Min S. Yun}
\affil{Dept.\ of Physics and Astronomy,
 University of Massachusetts, Amherst, MA 01003, USA}
\email{myun@astro.umass.edu}

\author{D. W. Hawkins}
\affil{Owens Valley Radio Observatory, California
 Institute of Technology, Big Pine, CA 93513, USA}
\email{dwh@ovro.caltech.edu}

\author{D. T. Frayer}
\affil{Spitzer Science Center, California Institute of
 Technology, Mail Code 220-06, Pasadena, CA 91125, USA}
\email{frayer@ipac.caltech.edu}

\and

\author{Kate G. Isaak}
\affil{Department of Physics and Astronomy, Cardiff
 University, 5 The Parade, Cardiff CF24~3YB UK}
\email{kate.isaak@astro.cf.ac.uk}

\begin{abstract}

Searches for CO emission in high-redshift objects have
traditionally suffered from the accuracy of
optically-derived redshifts due to lack of bandwidth in correlators 
at radio observatories.  This problem has motivated the creation
of the new COBRA continuum correlator, with 4~GHz available bandwidth,
at the Owens Valley Radio Observatory Millimeter Array.  Presented
here are the first scientific results from COBRA.  We report detections 
of redshifted CO($J=3\rightarrow2$) emission in the QSOs \smm  \ and 
VCV J140955.5+562827, as well as a probable detection in \rxj.  
At redshifts of $z=2.846$, $z=2.585$, and $z=2.796$, 
we find integrated CO flux densities of $5.4~\jykms$,
$2.4~\jykms$, and $2.9~\jykms$ for \smm, VCV J140955.5+562827, and 
\rxj, respectively, over linewidths  of $\Delta V_{FWHM} \sim 350~\kms$.
These measurements, when corrected for 
gravitational lensing, correspond to molecular gas masses of order 
$M(\hh) \sim 10^{9.6-11.1}\,\msun$, and are consistent with previous CO 
observations of high-redshift QSOs.  We also report $3\sigma$ upper limits on
CO(3$\rightarrow$2) emission in the QSO \lbqs \ of $1.3~\jykms$.  
We do not detect significant 3~mm continuum emission 
from any of the QSOs, with the exception of a tentative ($3\sigma$) 
detection in \rxj \ of $S_{3mm}=0.92~\mjybeam$.

\end{abstract}


\section{INTRODUCTION}

High-redshift QSOs have been shown to be often associated with 
dusty host galaxies (e.g.\ Barvainis \& Ivison 2002b), 
and as such offer much 
potential as powerful probes of the formation and evolution of 
massive galaxies in the universe.  The masses of high-redshift host
galaxies, as inferred by extrapolation from the locally 
measured correlation between central, supermassive QSO black 
hole masses and the velocity dispersion of stars in the host galaxy 
spheroids \citep{gms2000,fm2000} are large.  Thus, high-redshift
QSOs are signposts to the position and redshift of massive galaxies
in the early universe. Many high-redshift QSOs have been detected by recent 
millimeter and submillimeter surveys (e.g.\ Carilli et al.\ 2001; 
Isaak et al.\ 2002; Barvainis \& Ivison 2002b), implying that there 
is a link between QSOs and the massive, dusty, and gas-rich 
galaxy population discovered by deep submillimeter surveys 
(e.g.\ Smail, Ivison, \& Blain 1997; Cowie, Barger, \& Kneib 2002).
Most of these so-called submillimeter sources are too faint
for optical redshifts to be determined, even with the largest 
telescopes, precluding deep searches for the large molecular
gas reservoirs associated with young, massive galaxies. 
By studying large samples of dusty, high-$z$ QSOs
and the gas and dust properties of their host systems, we can 
evaluate their overlap with the submillimeter galaxy population 
and investigate their properties and evolution.

Recently, studies of CO emission in high-redshift objects have 
seen much success, with over 20 separate galaxies detected 
(e.g.\ Barvainis, Alloin, \& Bremer 2002a; Guilloteau et al.\ 1999),
the majority from galaxies with $S_{850\micron}>10$~mJy (e.g.\ 
Figure~1 in Isaak et al.\ 2002).  From these measurements it is
possible to infer the total molecular gas mass in the 
high-$z$ host galaxies, and in turn establish the fraction 
of the total mass that is still in gaseous form -- a 
suggestive indicator of evolutionary state.

CO measurements also provide long-sought evidence of 
previous metal enrichment, both in these objects 
and in the early universe, 
since carbon and oxygen are mainly 
produced by fusion reactions in stellar cores.  
These stars must also have produced copious amounts of other 
heavy elements.  We therefore expect to see metal-rich
gas ($Z \sim Z_{\sun}$) in massive starbursting systems 
very soon after the first starbursts (after 
$\sim 200$~Myr). The spectacular detections of both CO 
\citep{fabian_sdss,berCO}\ and inferred presence of dust 
\citep{berdust,sdssdust}\ in SDSS~1148+5251 
indicate that heavy-element enriched gas 
was already present in some objects at $z=6.4$, within 
800~Myr of the Big Bang.  Star formation must have
already been an important process by $z \sim 6.4$.  
The growing number of CO detections at lower, 
yet still cosmologically significant, redshifts,
are very important, indicating how widespread 
star forming systems were at early epochs, and supporting
the link between submillimeter galaxies and QSO hosts.

Searches for high-redshift CO have been hampered
by the limited spectrometer bandwidths (typically about 
500-1500 $\kms$) available to date on the current generation
of millimeter telescopes and interferometers.  
The lack of bandwidth prevents easy search for CO 
emission because the optically-determined redshifts for QSOs, 
typically measured from broad emission lines,
are known to be blueshifted from the host galaxy redshifts 
by as much as 1000-2000 $\kms$ \citep{tytlerfan}.  As an 
example, the QSO APM 08279+5255 has an optical redshift of 
$z=3.87$ from \citet{apmoptical}, determined through identification
of its rest-frame UV emission lines; by 
contrast, the redshift of the CO line of its host galaxy, 
from \citet{downes99}, is $z_{CO}=3.911$, a discrepancy 
of approximately 2500 $\kms$ (blueshifted).
At the Owens Valley Radio Observatory (OVRO), a new very 
broadband cross-correlator system, named COBRA, has been 
recently implemented to alleviate this problem.  
With a spectral coverage of 4 GHz, 
COBRA represents an eight-fold increase over the previous 
bandwidth at OVRO, and currently provides the largest 
available bandwidth for CO searches at millimeter wavelengths. 
This system is particularly advantageous for high-redshift 
spectral line searches when the redshift is poorly known 
\emph{a priori}.  Presented in Figure~1 is 
a sample COBRA spectrum, showing 3~GHz of the 4~GHz available bandwidth. 

As part of the correlator commissioning, we started a   
program to detect CO emission from high-redshift QSOs.  
To this end, we compiled a database of submillimeter-bright, 
$z>2$ QSOs from the literature from
which to choose a sample of targets.  To be included 
in the list, each object was required to meet the following
criteria: (1) detection in 2 or more submillimeter/far-infrared (FIR)
bands; (2) $T_{dust} \leq 100$ K, as derived following
\citet{MinSED}, such that the rest-FIR spectral 
energy distribution includes a significant cold 
dust contribution; and (3) FIR luminosity greater 
than or equal to that of the prototypical ULIRG Arp 220
(i.e.\ $L_{FIR}>10^{12.2}\,\lsun$).  We chose 
a total of 4 QSOs from the total of 23 sources in this database
to observe, selecting
those with the greatest FIR luminosities but without
a published CO detection: \lbqs, \smm, \rxj, and 
VCV J140955.5+562827.  In the sections that follow, 
we detail our observations and report the results 
of our efforts, which are part 
of an ongoing program of observations at OVRO.

Throughout this paper, we assume an 
$\Omega_{\textrm{M}}=0.3$, $\Omega_{\Lambda}=0.7$ cosmology 
with $H_{0}=70~\kms~\mpc^{-1}$.  We also
provide a parallel analysis in parentheses in the Einstein-deSitter 
cosmology $\Omega_{\textrm{M}}=1$ and $\Omega_{\Lambda}=0$,
with $H_{0}=75~\kms~\mpc^{-1}$.

\section{OBSERVATIONS AND DATA CALIBRATION}

A search for redshifted CO(3$\rightarrow$2)
($\nu_{rest}=345.796$~GHz) emission from the submillimeter-bright 
QSOs listed in Table~1 was made during 2003 May - June with 
the OVRO Millimeter Array in the compact (C) configuration
\footnote{In this configuration, the interferometer has baselines
ranging in length between 18 and 55~meters.} and the new COBRA correlator.
This is the lowest resolution configuration of the six 10.4-m 
telescopes, and is that most sensitive for searches for
millimeter emission from high-redshift sources.  
Listed in Table~1 are the observational parameters for each source.
The redshifts have been taken from the 
published broad optical emission line redshifts
with an offset applied in an attempt to correct for possible 
emission-line bias in the optical redshift, because previous 
searches near the broad-line redshift of some QSOs have
failed.  Typical single-sideband temperatures in the frequency
range 87-99~GHz were between 
200-400~K, corrected for antenna and atmospheric 
losses.  The weather during the observations was good,
especially for the time of year, and poor phase coherence
was rarely a problem.

The new COBRA digital cross-correlator provides 4~GHz 
of instantaneous bandwidth in each receiver sideband with 8 
32-channel spectrometers (each with 500 MHz BW). The correlator 
is based on FPGA chips operating at clock speeds of 125~MHz 
(Hawkins et al.\ in preparation).  At $\lambda \simeq 3$~mm, 
each 15.625 MHz-width channel corresponds to $\sim 50~\kms$ 
and so the full velocity coverage is 12,000 $\kms$ 
(c.f.\ galaxy CO linewidths, 100-1000~$\kms$,
velocity separations of interacting galaxies, $\lesssim 500~\kms$, and
galaxy cluster velocity dispersions, $\lesssim 1000~\kms$).  
For each observation, COBRA was
centered in the lower side-band at the frequency 
listed for each source in Table~1.  In addition, 3~mm
continuum data were recorded simultaneously with
the older, 2~GHz bandwidth analog continuum correlator at OVRO.

Nearby radio-loud quasars listed in Table~1 for each source were observed 
every 20 minutes for passband, gain, and phase calibration. 
Absolute flux calibration was determined from 
observations of various combinations of 3C\,273, 3C\,84, 
3C\,345, and 3C\,454.3, the flux histories of which are monitored by 
observations of Uranus and Neptune.  The absolute flux calibration 
uncertainty for the data is within 15\% for each source.  
The total integration time and spatial resolution for each source 
varied with system scheduling, source declination, and observed 
frequency, and is listed in Table~1.

The data were reduced using the latest version of the OVRO MMA software 
\citep{mmaref}, updated for use with COBRA data.  Data mapping 
was accomplished using standard tasks in MIRIAD \citep{miriadref}.

\section{RESULTS AND DISCUSSION}

\subsection{New Detections}

\subsubsection{\smm}

\smm, located behind the galaxy cluster Abell 478 ($z=0.088$), 
is the first type-1 QSO to have been discovered by nature of its 
submillimeter emission.  It was found in the Leiden-SCUBA Lensed 
Survey (Knudsen, van der Werf, \& Jaffe 2003) which is a survey 
of several galaxy cluster fields designed to detect gravitationally 
amplified, background submillimeter sources with the SCUBA 
instrument \citep{scuba}\ on the James Clerk Maxwell Telescope.  
One of the brightest submillimeter sources known, 
its fluxes measured with SCUBA are
$S_{850\micron}= 25 \pm 2.8$ mJy and $S_{450\micron}= 55 \pm 17$ mJy. 
The first optical-wavelength study of this 
QSO was recently published in \citet{opticalsmm}, in which a 
spectroscopic redshift of $z=2.837\pm 0.003$ from broad emission lines
as well as $I=20.5$ (not corrected for Galactic extinction; 
$I=19.4$ when a reddening of $E(\bv)=0.52$ is taken into account)
were measured.  The $I$-band source was found to be located 
$\sim 2 \arcsec$ southeast of the SCUBA position for \smm, 
within the source's SCUBA error circle\footnote{
The position uncertainty of SCUBA at $850 \micron$ is
$\pm 3\farcs2$.}.  The authors attribute the optical faintness 
of this QSO to a large viewing angle from the direction 
of relativistic beaming.  A gravitational lensing analysis by 
\citet{opticalsmm}\ using LENSTOOL \citep{lenstool}\ yields a 
magnification factor of 1.3.  The same authors, in an analysis of the 
galaxy's infrared (IR) SED (shown in Figure 3 of Knudsen et al. 2003), 
find $T_{dust}=29$~K, but say they cannot determine
the temperature of the hot dust (due to the AGN) because
of gaps in the IR SED.  Furthermore, the quasar is 
considered to be radio-quiet based on archived NVSS data, 
having a $3\sigma$ upper limit of 1.5~mJy at 1.4~GHz \citep{NVSS}.

A moment map of integrated CO(3$\rightarrow$2) emission 
for \smm \ is shown in Figure~2a.  No other significant 
CO sources at the same redshift are seen within the telescope's 
main beam, $\sim 70\arcsec$.  Obtaining the value of $\sigma$ 
for the map from the median single-channel rms per pixel over the 
64~pixel~$\times$~64~pixel map, excluding the 5 edge 
pixels on each side of the map, the peak of emission is detected 
at a significance level of $5.4\sigma$ at 89.911~GHz,   
corresponding to a redshift of $z_{CO}=2.846 \pm 0.002$. 
We calculate the uncertainty in the redshift according
to the formula \begin{equation}
\sigma_{z} = (\nu_{rest}/\nu_{obs}^2)(\Delta \nu_{obs})
\end{equation}
where $\nu_{rest}$ is the rest frequency of the line, 
$\nu_{obs}$ is the frequency at which the line is observed,
and $\Delta \nu_{obs}$ is the uncertainty in the center
position of the line.  
The value of $z_{CO}$ lies
slightly outside the stated error range of the broad-line 
redshift reported in \citet{opticalsmm}.  The discrepancy, 
$\Delta z=0.009$ or $\Delta v \sim +700~\kms$, is not surprising, 
and is in fact less
than the correction we applied to the optical redshift
($\Delta z = 0.018$) to search for CO.  By fitting 
a two-dimensional Gaussian profile directly 
to the emission map in the image plane, we find an upper 
limit to the source size of $14\farcs0 \times 7\farcs0$.  
The low resolution and low signal-to-noise (S/N) of our 
observations prevents us from determining any meaningful 
morphology of the QSO host from the map.  No 3~mm 
continuum emission was detected from \smm \ using data
from the 2~GHz bandwidth analog continuum correlator at OVRO,
with a 3$\sigma$ upper limit of $S_{3mm}<0.9\,\mjybeam$.  

In Figure~2a it can be seen that the peak CO emission from 
\smm \ is offset by several arcseconds to the northwest from 
the pointing (phase) center of the observation.  The 
coordinates published by \citet{opticalsmm}\ 
($\alpha_{J2000}=04^{h}13^{m}27\fs28$, 
$\delta_{J2000}=+10\degr27\arcmin41\farcs4$) explain this 
offset: they indicate that the position we observed was too 
far to the east.  The direction of offset from the SCUBA position 
is the same as that found in \citet{opticalsmm}.  However, the 
low S/N ratio of our observations means that
the absolute astrometry of our observations is 
dominated entirely by phase noise and baseline errors; 
thus, coordinates with absolute astrometry superior to those 
describing the optical position cannot be meaningfully derived.

By fitting a Gaussian profile to the observed spectrum,
which is taken from the location of the
peak emission in the image plane, not from the phase
center of the map,
we find that the velocity width (FWHM) of the CO(3$\rightarrow$2) 
line of \smm \ is $340 \pm 120~\kms$.  
We consider this width uncertain, however, due to a low S/N
ratio in the spectral amplitude.  The uncertainty in the linewidth
was found by plotting Gaussian profiles of varying FWHM
over the observed spectrum and choosing which profiles
could reasonably be said to fit the line.  The integrated flux density
over the width of this line,
found by adding the flux in the map pixels (excluding any pixels with
flux density less than the $1 \sigma$ value found for the map)
in a box centered on, and surrounding, the peak emission and dividing the
total by the beam area in pixels, 
is $5.4 \pm 1.3~\jykms$.  We calculate the uncertainty in the integrated flux, 
$S_{CO} \Delta V$, using the formula \begin{equation}
\sigma(S_{CO} \Delta V) = (\sigma_{chan})(\Delta V_{chan})(\sqrt{N_{chan}})(\sqrt{N_{pix,box}/N_{pix,beam}})
\end{equation}
where $\sigma_{chan}$ is the median single-channel rms per pixel, 
$\Delta V_{chan}$ is 
the velocity width of a channel, $N_{chan}$ is the number of channels
integrated, $N_{pix,box}$ is the number of pixels in the box of 
integration, and $N_{pix, beam}$ is the beam area in pixels.  

Using the relations for $L_{CO}$ and $L'_{CO}$ given in
Solomon, Downes, \& Radford (1992b),
and applying the lensing amplification 
correction factor of 1.3 \citep{opticalsmm}\, we obtain for \smm \ 
$L_{CO}=2.2\times10^{8}\,\lsun$ ($8.8\times10^{7}\,\lsun$)
and $L^{\prime}_{CO}=1.7\times10^{11}\,\Kkmspc$ 
($8.7\times10^{10}\,\Kkmspc$).  
Again using \citet{sdreqpaper}, with a $\hh$ to CO 
conversion factor of $\alpha=0.8\,\msun~(\Kkmspc)^{-1}$, 
which is the conversion factor \citet{dulirg}\ find for ULIRGs,
we infer a molecular gas mass for \smm \ of 
$M(\hh)=1.3\times10^{11}\,\msun$
($5.3\times10^{10}\,\msun$).  This galaxy is one of 
the most massive CO systems known to date.

\subsubsection{VCV J140955.5+562827}

VCV J140955.5+562827, hereafter referred to as J1409+5628, 
is a broad-absorption line (BAL) QSO, found in the Second 
Byurakan Survey (identifier SBS 1408+567) \citep{jrussjour}.  
It is significantly brighter optically than \smm, with $V=17.20$ 
and $\bv=0.40$, as measured by \citet{jnewphot}.  \citet{bal93}\ 
derive a redshift of $z=2.562$ by cross-correlating the QSO's
broad emission line spectrum with a template spectrum (formed 
by de-redshifting a large sample of BAL spectra by-eye, subtracting 
a low-order polynomial fit to emission-line free regions,
then combining all the spectra to form one high S/N spectrum).  
J1409+5628 is also a 2MASS point source, with
$J=15.95$, $H=15.16$, and $K=14.86$ \citep{2mass}.
The QSO has recently been detected in 1.2 mm continuum emission 
by \citet{omont1mm}, with $S_{250GHz}=10.7 \pm 0.6$~mJy.  
\citet{omont1mm} state that it is one of seven high-$z$
sources with $S_{250GHz}\gtrsim 10$~mJy.  There is, however,  
no indication that the object is strongly lensed, which suggests
that the QSO is intrinsically very luminous.  A search of the
NVSS archive suggests that J1409+5628 is radio-quiet, with a $3\sigma$ 
upper limit of 1.5~mJy at 1.4~GHz \citep{NVSS}.

Calculating all of our uncertainties in the same way as
described in \S 3.1.1, 
we detect CO(3$\rightarrow$2) emission with significance 
$5.0\sigma$ at 96.462~GHz, corresponding to a redshift of 
$z_{CO}=2.585 \pm 0.001$.  The spectrum of J1409+5628, taken
from the phase center of the map,
is shown in Figure~1.  By fitting a simple Gaussian line
profile, we derive a velocity width (FWHM) of
$370 \pm 60~\kms$. Figure~3 shows a close-up view of the
single COBRA band in which the line is detected, with
the fitted line profile displayed.

The map of integrated CO(3$\rightarrow$2) emission for J1409+5628 
is shown in Figure~2b.  The peak of the integrated emission
occurs at the phase center of our map, with no other CO sources
seen at the same redshift within the primary beam of the telescope. 
The integrated flux density over the width of the CO line is 
$2.4 \pm 0.7~\jykms$.  Again using the relations for $L_{CO}$ and 
$L'_{CO}$ from \citet{sdreqpaper}\ we obtain for J1409+5628  
$L_{CO}=1.1\times10^{8}\,\lsun$ ($4.4\times10^{7}\,\lsun$) and 
$L^{\prime}_{CO}=8.2\times10^{10}\,\Kkmspc$
($3.3\times10^{10}\,\Kkmspc$).
We infer a molecular gas mass for J1409+5628 of 
$M(\hh)=6.6\times10^{10}\,\msun$ ($2.7\times10^{10}\,\msun$).
From a direct Gaussian fit to the image plane of the integrated 
CO map, we place an upper limit to the source size of 
$6\farcs3 \times 11\farcs1$.  No significant 3~mm continuum 
emission is detected from this object, rather we determine a 
$3\sigma$ upper limit of $S_{3mm}<0.9\,\mjybeam$.

We note that the CO redshift for J1409+5628 is redshifted 
from the optical, broad-line redshift by $\Delta z=0.023$.  The 
corresponding discrepancy in velocity space is 
$\Delta v \sim +1900~\kms$.

\subsubsection{\rxj}

\rxj \ is a mini-BAL QSO\footnote{
The mini-BAL designation indicates that the velocity spread of 
the QSO's absorption trough, 3200~$\kms$, is just outside the 
usual velocity range that defines a BAL QSO.},
discovered through optical 
identification of ROSAT All-Sky Survey sources \citep{rxjdisc}. 
The source is gravitationally lensed by an elongated galaxy at $z=0.769$ 
(Kneib, Cohen, \& Hjorth 2000), which produces the 4 images in
an unusual configuration in optical images \citep{burud}.  
The appearance of the images is thought to be due to a large external
shear caused by a massive galaxy cluster, also at $z=0.769$,
at a projected distance of $38\arcsec$ \citep{burud,kneib}. 
The QSO redshift is $z=2.800$, found by fitting Gaussian
profiles to the neutral hydrogen Lyman-$\alpha$ and  
the \ion{C}{4} ($\lambda=1549$~\AA) lines of each QSO image, then
taking the average of the resulting redshift of each image \citep{rxjdisc}. 
\citet{rxjsubmm}\ measure large submillimeter flux densities of 
$S_{850\micron}=26.7 \pm 1.4$~mJy and $S_{450\micron}=65 \pm 19$~mJy, 
and infer from these values that the host galaxy is likely to have a 
significant cold dust component.  \citet{barCOsearch}\ 
attempted, without success, to detect \rxj \ in CO(3$\rightarrow$2)
emission with the IRAM Plateau de Bure interferometer, at the 
same redshift we adopt in the current observations ($z=2.807$),
but using a more extended configuration and so with higher 
spatial resolution.  They attribute the lack of detection to 
the combination of uncertainty in the optical redshift and 
small instrumental bandwidth (560 or 595~MHz, not specifically
stated for each source), finding a line
rms of 1.2~mJy/100~$\kms$.  \citet{barCOsearch}\ 
also report 1 and 3~mm continuum detections for \rxj \ of 
$S_{1mm}=10.2 \pm 1.8\,\mjybeam$ and 
$S_{3mm}=1.7 \pm 0.3\,\mjybeam$, which, together with the 
SCUBA data, fit well the
scenario that \rxj \ has a significant cold dust component.  
We note that \rxj \ is 
radio-quiet: the FIRST Survey \citep{first}\ reports a 
1.4~GHz flux density upper limit of 0.45~mJy. 

We tentatively detect CO(3$\rightarrow$2) emission from \rxj \ 
with a significance of $3.7\sigma$, again calculating
all uncertainties as in \S 3.1.1.  The line is centered at 
91.088~GHz, which corresponds to a redshift of 
$z_{CO}=2.796 \pm 0.001$.  We note that 
the CO redshift of \rxj \ turns out to be slightly blueshifted 
relative to the broad-line redshift, 
$\Delta v \sim -320~\kms$, and places the line
emission peak just outside the bandwidth
of the correlator used by \citet{barCOsearch}.  No other 
significant CO sources are seen within the array's primary
beam; the resolution of our observations is not sufficient 
to resolve the multiple images of this QSO that one would expect
from the lensing model.  The integrated CO emission map is 
shown in Figure~2c, where we see that the peak of
the emission is slightly north of the phase center of the 
observations. 
From a direct Gaussian fit to this map, we find an upper 
limit to the source size of $13\farcs0 \times 6\farcs9$.  The 
velocity width (FWHM) of the CO line is $350 \pm 60~\kms$, 
and the integrated flux density over this line width is 
$2.9 \pm 1.1~\jykms$.  Uncorrected for lensing, the integrated 
flux corresponds to a CO luminosity 
and molecular gas mass of $L_{CO}=1.5\times10^{8}\,\lsun$
($6.0\times10^{7}\,\lsun$), 
$L^{\prime}_{CO}=1.1\times10^{11}\,\Kkmspc$
($4.5\times10^{10}\,\Kkmspc$), 
and $M(\hh)=9.1\times10^{10}\,\msun$
($3.6\times10^{10}\,\msun$), 
from the relations in \citet{sdreqpaper}.  Using the 
lens magnification factor of 21.8 quoted in \citet{rxjsubmm},
these quantities decrease to the following: 
$L_{CO}=6.9\times10^{6}\,\lsun$ ($2.8\times10^{6}\,\lsun$), 
$L^{\prime}_{CO}=5.2\times10^{9}\,\Kkmspc$ ($2.1\times10^{9}\,\Kkmspc$), 
and $M(\hh)=4.2\times10^{9}\,\msun$
($1.7\times10^{9}\,\msun$). 

We tentatively detect 3~mm continuum emission from \rxj,
identifying a $\sim 3\sigma$ source at the phase center of the 
continuum map with flux density $S_{3mm}=0.92 \pm 0.29\,\mjybeam$.  
This value for $S_{3mm}$ differs from \citet{barCOsearch} by
$0.8~\mjybeam$, thus the values do not formally agree.
Our tentative detection may be spurious, however, 
due to phase offsets and other errors collecting at the phase 
center of a map.

\subsection{Non-Detection: \lbqs}

\lbqs \ is a QSO found in the Large, Bright Quasar Survey 
(LBQS) \citep{lbqsorig}.  \citet{lbqsz}\ find a broad-line 
redshift of $z=2.596 \pm 0.005$, though we searched for CO 
at the original redshift measurement from \citet{lbqsorig}, 
$z=2.56$.  The LBQS measures a photographic magnitude of 
$B=17.4$, but more recently, \citet{lbqsmag}\ report CCD 
magnitudes of $g^{*}=17.20$ and $r^{*}=16.95$. \lbqs \ is 
also detected by 2MASS \citep{2mass}, with measured near-infrared 
magnitudes of $J=16.2$ and $H=15.4$.  \lbqs \ has a narrow absorption 
line on its Lyman-$\alpha$ line, according to \citet{lbqsnal}.  
\citet{lbqssubmm}\ report a submillimeter flux density of 
$S_{850\micron}=17.2 \pm 2.9$~mJy measured with SCUBA. No 
radio continuum emission has been observed from this QSO, 
with $3\sigma$ upper limits on flux density of 0.18~mJy at 
8.4~GHz \citep{lbqsradio}\ and 1.5~mJy at 1.4~GHz \citep{NVSS}.

We do not detect CO(3$\rightarrow$2) emission from \lbqs, 
despite the 4~GHz bandwidth made available by the COBRA correlator, 
which corresponds to a range of redshift of $2.489<z<2.631$.  
Our observations yield a single-channel rms of 3.3~$\mjybeam$, 
which in turn corresponds to a $3\sigma$-upper limit to the 
integrated flux density at 97.1~GHz of 1.3~$\jykms$, assuming 
a typical line FWHM of 300~$\kms$ and using equation (2) in
\S 3.1.1.  This integrated flux places 
upper limits to the CO luminosity and molecular gas mass of 
$L_{CO}<5.8\times10^{7}\,\lsun$ ($2.4\times10^{7}\,\lsun$), 
$L^{\prime}_{CO}<4.4\times10^{10}\,\Kkmspc$ 
($1.8\times10^{10}\,\Kkmspc$), and
$M(\hh)<3.5\times10^{10}\,\msun$ ($1.4\times10^{10}\,\msun$). 

As this work was in revision, it was made known to us that 
CO(3$\rightarrow$2) emission from \lbqs \ had been detected 
with the IRAM Plateau de Bure interferometer at the same time,
at a level just below the sensitivity limit of our OVRO 
observations of this object.

\subsection{High-z CO Emitters as a Whole}

In Table~2, we have summarized the $z>1$ objects with CO 
detections published to date, including the results presented 
here.  At least 25 high-redshift objects are known CO emitters, 
with a number of objects detected in more than 
one CO transition.  At the bottom of the table, the mean and median
values of $\Delta V_{FWHM}$ and $S_{CO} \Delta V$ are given, 
for the sample of CO(3$\rightarrow$2) detections only as well
as the sample of all the objects in the table. 

The three detections reported in this paper, 
for J1409+5628, \rxj, and \smm, have velocity widths
and integrated line fluxes that lie within the range set 
by the $z>1$ sample of CO detections.  With an integrated
flux nearly twice the mean of the CO(3$\rightarrow$2) 
sample, \smm \ may have the 
largest lensing-corrected molecular gas mass of any galaxy yet detected, 
even greater than that of PSS~2322+1944 
(with $M(\hh)=5.5\times 10^{10}\,\msun$ after using the lensing
magnification factor 2.5 from Carilli et al.\ 2003).  
The low dust temperature inferred by \citet{opticalsmm}\ 
suggests that not all of the 
submillimeter emission can be produced by an AGN, and together 
with the high molecular gas mass, it suggests that 
this galaxy is undergoing a massive burst of star formation.
By contrast, the measurements of integrated flux for J1409+5628 
and \rxj \ lay close to the median and mean of the
CO($3\rightarrow2$) detection sample, respectively.

Figure~4 shows a histogram of the redshifts of the objects listed 
in Table~2, with a bin size of $\Delta z = 0.2$.  
This diagram represents a graphical summary of the status of 
the field of high-$z$ CO detection, and is not intended
to indicate any statistical trends.  Many of the reported 
detections cluster around $z\sim 2.5$, and one may
point out that this clustering appears similar to the 
distribution of redshifts of submillimeter galaxies 
(median$(z)=2.4$, Chapman et al.\ 2003).  However, 
that CO detections are most numerous here may not be surprising, 
given that the most easily observed millimeter band
using currrent interferometers is the 3~mm band  (85 to 115~GHz), 
and this band corresponds to CO(3$\rightarrow$2) at $z \sim 2-3$.  

A sample size of just 25 CO emitters is too small to draw any
statistically significant conclusions, given the range of
redshift and the selection criteria that have been used to
identify the sources.  Nonetheless, we see from 
Figure~4 that CO emission can be detected over the entire range 
of redshift currently observable by available instruments.  From 
the standpoint of galaxy and QSO co-evolution, this prompts the 
question of whether CO, and by inference large masses of gas with
significant heavy element enrichment, can be detected in all IR-bright quasars, 
particularly those at high redshift?  Of optically selected,
IRAS detected, nearby, PG QSOs, a large fraction have CO detections 
\citep{pg1,pg2,pg3}.  

At high redshift, however, the fraction of known QSOs 
with CO detections is not high, likely due to 
limited ranges of redshift that are searched (limited in 
spectrometer bandwidth and target selection) and large required 
integration times. However, the growing number of submillimeter
QSO detections suggests that there may yet be many more high-$z$ 
CO sources, since submillimeter flux density is proportional to 
the intrinsic FIR luminosity and ISM mass of a galaxy.  
We mention in \S 1 that the majority of high-$z$ CO detections 
come from galaxies with $S_{850\micron}>10$~mJy.  In fact, the 
10 brightest submillimeter QSOs with measured redshifts, all 
with $S_{850\micron}>20$~mJy, are CO sources (B1938+666, with 
$S_{850\micron}=35$~mJy from Barvainis \& Ivison 2002b, has no 
published redshift).  30\% of high-$z$ QSOs are detected in 
the millimeter and/or submillimeter continuum and most of those 
galaxies have CO detections (see Omont et al.\ 2001; 
Omont et al.\ 2003; Priddey et al.\ 2003a; Priddey et al.\ 2003b;
Isaak et al.\ 2002).  There are at least 
20 more QSOs with $z>1$ in the literature with recently published values 
of $S_{850\micron}$ in the range 8-10~mJy in which a CO detection 
has not been published.  If these QSOs are CO
sources, not only would they increase significantly the fraction of 
high-$z$ QSOs with CO detections, but also 
strengthen the link between the submillimeter galaxy population 
and the QSO phenomenon and support the scenario of co-evolution 
of galaxies and massive black holes.  

The relative luminosity contributions from star formation and AGN 
to the large submillimeter emission of high-$z$ QSOs is a related 
issue of much interest.  \citet{minnew}\ suggest that, in the 
nearby universe, optically selected QSOs are distinct in nature
from the ULIRG population even if their host galaxies are rich 
in molecular gas.  They suggest that the presence of 
large molecular gas reservoirs in QSOs does not necessarily 
imply a high rate of star formation; thus, the fact that we detect
CO in \smm, J1409+5628, and \rxj \ does not 
definitively show that these objects' luminosities are
dominated by star formation.  We must seek avenues other
than CO detection to understand the sources of a galaxy's luminosity.
In the case of high-$z$ QSOs, the galaxies are too distant to
determine independently if the QSO hosts are undergoing 
starbursts (e.g.\ by detecting clumpy, irregular morphology,
or by imaging young, super star clusters in the host galaxy).
However, the detection of the HCN(1$\rightarrow$0) transition in the 
submillimeter-bright, 
CO detected, H1413+117 (the Cloverleaf quasar, $z=2.558$) by 
\citet{cloverhcn}\ suggests that at least some high-$z$ IR-bright 
QSOs experience major star formation activity during their QSO phase, 
and that a significant part of the FIR emission comes from the star 
formation (as opposed to coming from the AGN).  To determine if our, 
and other, submillimeter-bright, molecular gas-rich,
high-$z$ QSOs have a high percentage of their FIR luminosity 
due to star formation, these galaxies should also be searched 
for HCN emission.  However, it may be some time before
such an investigation becomes practical, as the recent work of 
both \citet{cloverhcn}\ and \citet{brhcn}\ imply that only the 
most luminous HCN sources can be detected at high redshift with   
currently available instruments.  Isaak et al.\ (2004) suggest 
that we will have to wait until the EVLA\footnote{Details of the
EVLA may be found at http://www.aoc.nrao.edu/evla/} comes on line
in order to have the necessary sensitivity to detect HCN in many 
high-$z$ galaxies.

Finally, while a thorough, systematic evaluation of the range and sense of 
the relative redshift differences between the QSO and its host galaxy
is beyond the scope of this paper, we wish to draw attention to the
variation in this offset in Figure~5, which is a plot of the 
difference between the published optical redshift (representing
the QSO redshift) and the CO redshift (representing the host galaxy
redshift) in the sense $z_{CO}-z_{optical}$.   
The need for the large bandwidth of the COBRA correlator 
at OVRO is apparent in Figure~5, since the range of
redshift offsets is large, and it is impossible to know 
\emph{a priori} how different the host galaxy's redshift will be.  The 
clustering of points (representing CO-detected objects only)
about the $z_{CO}-z_{optical} = 0$ 
line and the relatively small number of outliers likely reflects 
the bias of the narrow bandwidths used in the past, toward detecting 
objects with small redshift offsets.  Objects searched for CO, 
but not detected due to unknown
redshift offsets, are not included in the plot.  Undoubtedly
such cases exist and will increase the dispersion in Figure~5.  
The large increase in spectral bandwidth provided by COBRA will certainly
increase the likelihood of actually having a galaxy's CO line 
fall in the observed spectral window.

\section{CONCLUSIONS}

Three more high-$z$ QSOs have been detected in CO emission, \smm \ 
at $z=2.846$, VCV J140955.5+562827 at $z=2.585$, and \rxj \ at $z=2.796$,
using the new COBRA correlator at the Owens Valley Radio Observatory.  
All three submillimeter-bright QSOs possess 
large reservoirs of molecular gas, further linking submillimeter 
galaxies to their QSO contemporaries, and
\smm \ may be one of the most massive CO systems known.

\acknowledgments

We thank the anonymous referee for helpful comments to improve this
paper.  We also thank the staff of the Owens Valley Radio Observatory.
The Owens Valley Millimeter Array is supported by National Science 
Foundation (NSF) grant AST~99-81546.   LJH acknowledges support from an NSF Graduate 
Research Fellowship.  This research has made use of the NASA/IPAC 
Extragalactic Database (NED) which is operated by the Jet Propulsion 
Laboratory, California Institute of Technology, under contract with 
the National Aeronautics and Space Administration.

\clearpage

\begin{deluxetable}{lccccccccl}
\tabletypesize{\tiny}
\tablewidth{0pt}
\tablenum{1}
\tablecolumns{10}
\tablecaption{Observational Parameters of Sub-mm Bright QSOs}
\tablehead{
\colhead{Source} & \colhead{R.A.\tablenotemark{a}} & \colhead{Declination\tablenotemark{a}} & 
\colhead{Redshift} & \colhead{$M_{B}$} & 
\colhead{$S_{850 \micron}$} & \colhead{Frequency} & \colhead{Beam Size\tablenotemark{b}} & 
\colhead{$t_{int}$\tablenotemark{c}} & \colhead{Calibrator} \\
\colhead{} & \colhead{(J2000.0)} & \colhead{(J2000.0)} & \colhead{} & \colhead{} &
\colhead{(mJy)} & \colhead{(GHz)} & \colhead{( $\arcsec \times \arcsec$ )} & 
\colhead{(hours)} & \colhead{}}

\startdata
\lbqs & 00\phm{:}21\phm{:}27.30 & $-$02\phm{:}03\phm{:}33.0 & 2.560 & $-$28.6 & 17 & 97.13 & $13 \times 11$ & 10 & 0006$-$063 \\
\smm & 04\phm{:}13\phm{:}27.50 & $+$10\phm{:}27\phm{:}40.3 & 2.855 & \nodata & 25 & 89.70 & $15 \times 11$ & 26 & 0433+053 \\
\rxj & 09\phm{:}11\phm{:}27.50 & $+$05\phm{:}50\phm{:}52.0 & 2.807 & \nodata & 27 & 90.83 & $14 \times 10$ & 29 & 0854+201 \\
VCV J140955.5+562827 & 14\phm{:}09\phm{:}55.50 & $+$56\phm{:}28\phm{:}27.0 & 2.560 & $-$28.4 & 11\tablenotemark{d} & 97.13 & $14 \times 10$ & 28 & 1419+543 \\
\enddata
\tablecomments{Units of right ascension are hours, minutes, and seconds.  Units of declination
are degrees, arcminutes, and arcseconds.}
\tablenotetext{a}{Optical QSO position.}
\tablenotetext{b}{Theoretical FWHM of OVRO beam at source declination and observed frequency, 
assuming natural weighting.}
\tablenotetext{c}{Total effective on-source integration time with six telescopes.}
\tablenotetext{d}{1.2 mm flux density, as $850 \micron$ flux density has not been measured.}
\end{deluxetable}

\clearpage
\begin{deluxetable}{lccccccc}
\tabletypesize{\tiny}
\tablewidth{0pt}
\tablenum{2}
\tablecolumns{8}
\tablecaption{High-z CO Detections Published To Date}
\tablehead{
\colhead{Source} & \colhead{Type\tablenotemark{a}} & \colhead{$z_{CO}$} & \colhead{$S_{850 \micron}$} &
\colhead{Transition} & \colhead{$\Delta V_{FWHM}$} & \colhead{$S_{CO}$$\Delta V$} & 
\colhead{Reference} \\
\colhead{} & \colhead{} & \colhead{} & \colhead{(mJy)} & \colhead{} &\colhead{($\kms$)} & 
\colhead{($\jykms$)} & \colhead{}}

\startdata
0957+561 & QSO & 1.414 & 7.5 & CO(2$\rightarrow$1) & 440 & 1.2 & 1,2 \\
HR~10 & ERO & 1.439 & 4.9 & CO(2$\rightarrow$1) & 400 & 1.4 & 3,4 \\
F~10214+4724 & QSO/ULIRG & 2.285 & 50 & CO(3$\rightarrow$2) & 230 & 4.8 & 5,6 \\
SMM~J16358+4057 & Submm & 2.385 & 8.2 & CO(3$\rightarrow$2) & 840 & 2.3 & 7,8 \\
53W002 & RG & 2.394 & 3.1 & CO(3$\rightarrow$2) & 540 & 1.5 & 9,10 \\
SMM~J04431+0210 & Submm & 2.509 & 7.2 & CO(3$\rightarrow$2) & 350 & 1.4 & 7,11 \\
H~1413+117 & QSO & 2.558 & 66 & CO(3$\rightarrow$2) & 330 & 9.9 & 12,2 \\
SMM~J14011+0252 & Submm & 2.565 & 6 & CO(3$\rightarrow$2) & 200 & 2.4 & 13,14 \\
\textbf{VCV J1409+5628} & \textbf{QSO} & \textbf{2.585} & \nodata & \textbf{CO(3$\rightarrow$2)} &
 \textbf{370} & \textbf{2.4} & \textbf{(this paper)} \\
MG~0414+0534 & QSO & 2.639 & 25 & CO(3$\rightarrow$2) & 580 & 2.6 & 15,2 \\
MS~1512-cB58 & LyB & 2.727 & 4.2 & CO(3$\rightarrow$2) & 170 & 0.4 & 16 \\
LBQS~1230+1627B & QSO & 2.735 & 29 & CO(3$\rightarrow$2) & \nodata & 0.8 & 17 \\
\textbf{\rxj} &  \textbf{QSO} & \textbf{2.796} & \textbf{27} & \textbf{CO(3$\rightarrow$2)} &
 \textbf{350} & \textbf{2.9} & \textbf{(this paper)} \\
SMM~J02399-0136 & QSO/Submm & 2.808 & 25 & CO(3$\rightarrow$2) & 710 & 3.0 & 18,19 \\
\textbf{\smm} & \textbf{QSO/Submm} & \textbf{2.846} & \textbf{25} & \textbf{CO(3$\rightarrow$2)} &
 \textbf{340} & \textbf{5.4} & \textbf{(this paper)} \\
B3~J2330+3927\tablenotemark{b} & RG & 3.094 & 14 & CO(4$\rightarrow$3) & 220 & 1.3 & 31 \\
MG~0751+2716 & QSO & 3.208 & 26 & CO(4$\rightarrow$3) & 390 & 6.0 & 20,2 \\
SMM~J09431+4700 & Submm & 3.346 & 10.5 & CO(4$\rightarrow$3) & 420 & 1.1 & 7,21 \\
TN~J0121+1320\tablenotemark{b} & RG & 3.520 & \nodata & CO(4$\rightarrow$3) & 300 & 1.2 & 32 \\ 
6C1909+722 & RG & 3.532 & 14 & CO(4$\rightarrow$3) & 530 & 1.6 & 22 \\
4C\,60.07 & RG & 3.791 & 11 & CO(4$\rightarrow$3) & 550 & 2.5 & 22 \\
APM~08279+5255 & QSO & 3.911 & 75 & CO(4$\rightarrow$3) & 400 & 3.7 & 23,24 \\
PSS~2322+1944 & QSO & 4.120 & 23 & CO(4$\rightarrow$3) & 375 & 4.2 & 25,26 \\
BR~1335-0414 & QSO & 4.407 & 14 & CO(5$\rightarrow$4) & 420 & 2.8 & 27,28 \\
BR~0952-0115 & QSO & 4.434 & 13 & CO(5$\rightarrow$4) & 230 & 0.9 & 17,2 \\
BR~1202-0725 & QSO & 4.690 & 42 & CO(5$\rightarrow$4) & 320 & 2.4 & 29,28 \\
SDSS~1148+5251 & QSO & 6.419 & \nodata & CO(3$\rightarrow$2) & 380 & 0.2 & 30 \\[5pt]
\textbf{Mean CO(3$\rightarrow$2)} & & & & & \textbf{415} & \textbf{2.9} &  \\
\textbf{Median CO(3$\rightarrow$2)} & & & & & \textbf{350} & \textbf{2.4} & \\[3pt]
\textbf{Mean (all)} & & & & & \textbf{399} & \textbf{2.6} &  \\
\textbf{Median (all)} & & & & & \textbf{380} & \textbf{2.4} & \\

\enddata

\tablenotetext{a}{Key to abbreviations -- QSO = quasi-stellar object; ERO = extremely red object; 
ULIRG = ultraluminous infrared galaxy; Submm = submillimeter galaxy; 
RG = radio galaxy; LyB = Lyman break galaxy} 

\tablenotetext{b}{Added in proofs.}

\tablerefs{(1) Planesas et al.\ 1999; (2) Barvainis \& Ivison 2002; (3) Andreani et al.\ 2000; (4) Dey et al.\ 1999; 
(5) Solomon, Downes, \& Radford 1992a and Downes, Solomon, \& Radford 1995; (6) Rowan--Robinson et al.\ 1993; 
(7) Neri et al.\ 2003; (8) Ivison et al.\ 2002; (9) Scoville et al.\ 1997; (10) Smail et al.\ 2003; 
(11) Smail et al.\ 1997 and Smail et al.\ 2002; (12) Barvainis et al.\ 1994; (13) Frayer et al.\ 1999; 
(14) Smail et al.\ 1998; (15) Barvainis et al.\ 1998; (16) Baker et al.\ 2004; (17) Guilloteau 
et al.\ 1999; (18) Frayer et al.\ 1998; (19) Ivison et al.\ 1998; (20) Barvainis et al.\ 2002a;
(21) Cowie et al.\ 2002; (22) Papadopoulos et al.\ 2000; (23) Downes et al.\ 1999; 
(24) Lewis et al.\ 1998; (25) Cox et al.\ 2002; (26) Isaak et al.\ 2002; (27) Guilloteau et al.\ 1997; 
(28) quoted in McMahon et al.\ 1999; (29) Omont et al.\ 1996; (30) Walter et al.\ 2003; (31) De Breuck et al.\ 2003a;
(32) De Breuck, Neri, \& Omont 2003b}

\end{deluxetable}

\clearpage

\begin{figure}
\figurenum{1}
\plotone{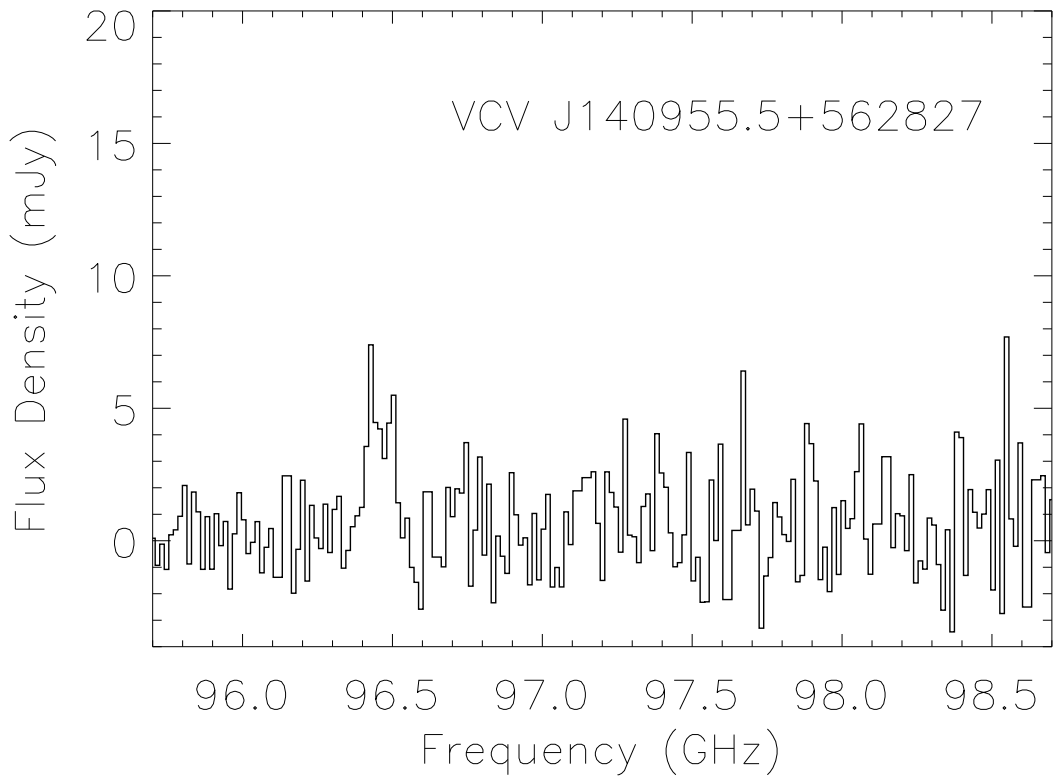}
\caption{Example spectrum from COBRA correlator, showing 3~GHz of 
the available 4~GHz bandwidth.}
\end{figure}

\begin{figure*}
\figurenum{2}
\includegraphics*[width=6.0 cm, angle=-90]{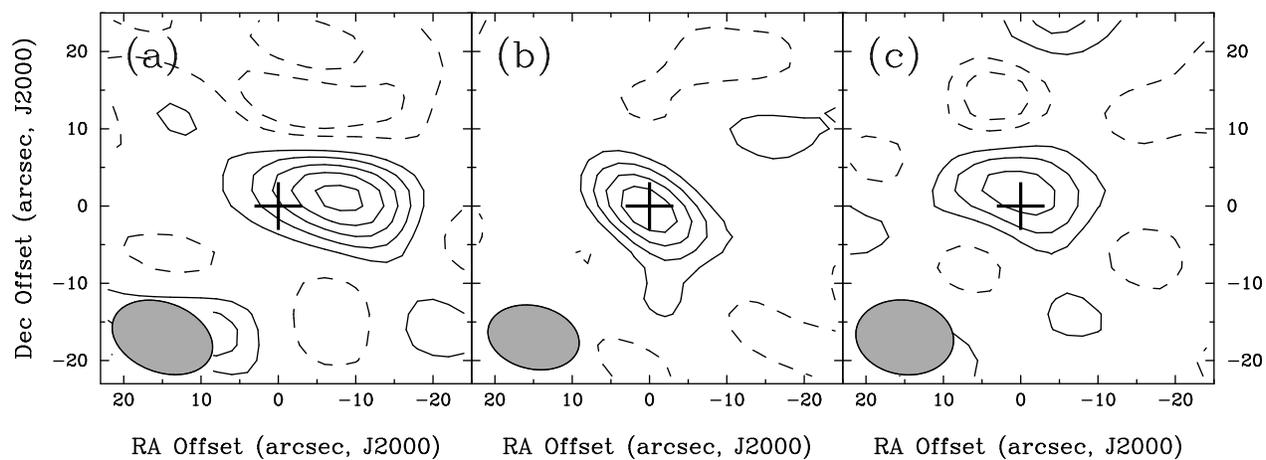}
\caption{Maps of integrated CO emission in (a) \smm, (b) J1409+5628,
and (c) \rxj. The contours are multiples (-2,-1,1,2,3,4,5) of the rms noise
levels, which are 0.92~$\jybeam\,\kms$, 0.49~$\jybeam\,\kms$,
and 0.88~$\jybeam\,\kms$.  The crosses mark the phase center in each
map.  North is up, west is to the right.}
\end{figure*}

\begin{figure}
\figurenum{3}
\plotone{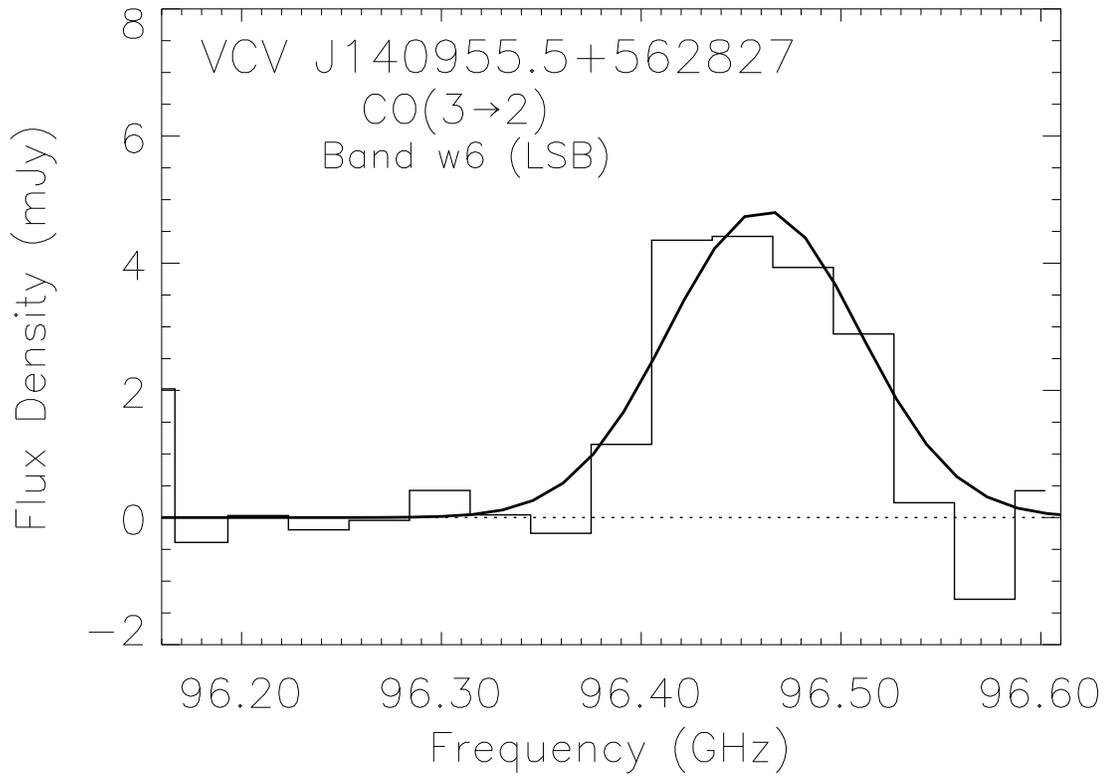}
\caption{Spectral line fit for J1409+5628.  The spectrum has been 
continuum-subtracted and smoothed to a resolution of 97~$\kms$.}
\end{figure}

\begin{figure}
\figurenum{4}
\plotone{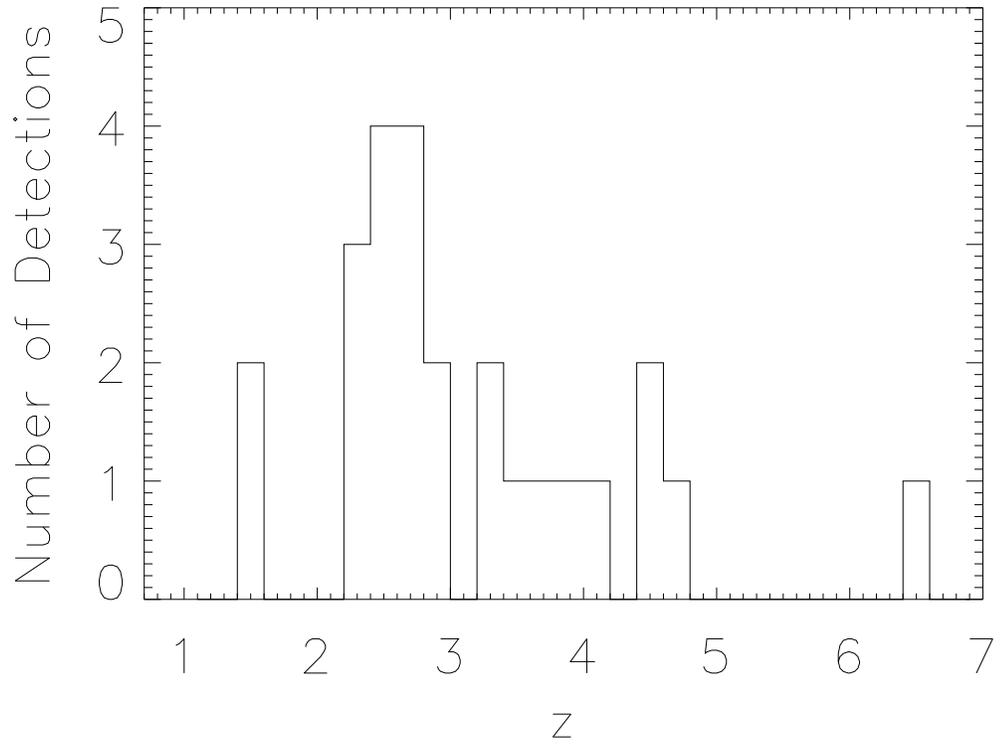}
\caption{Histogram of redshifts of high-z CO detections.}
\end{figure}

\begin{figure}
\figurenum{5}
\plotone{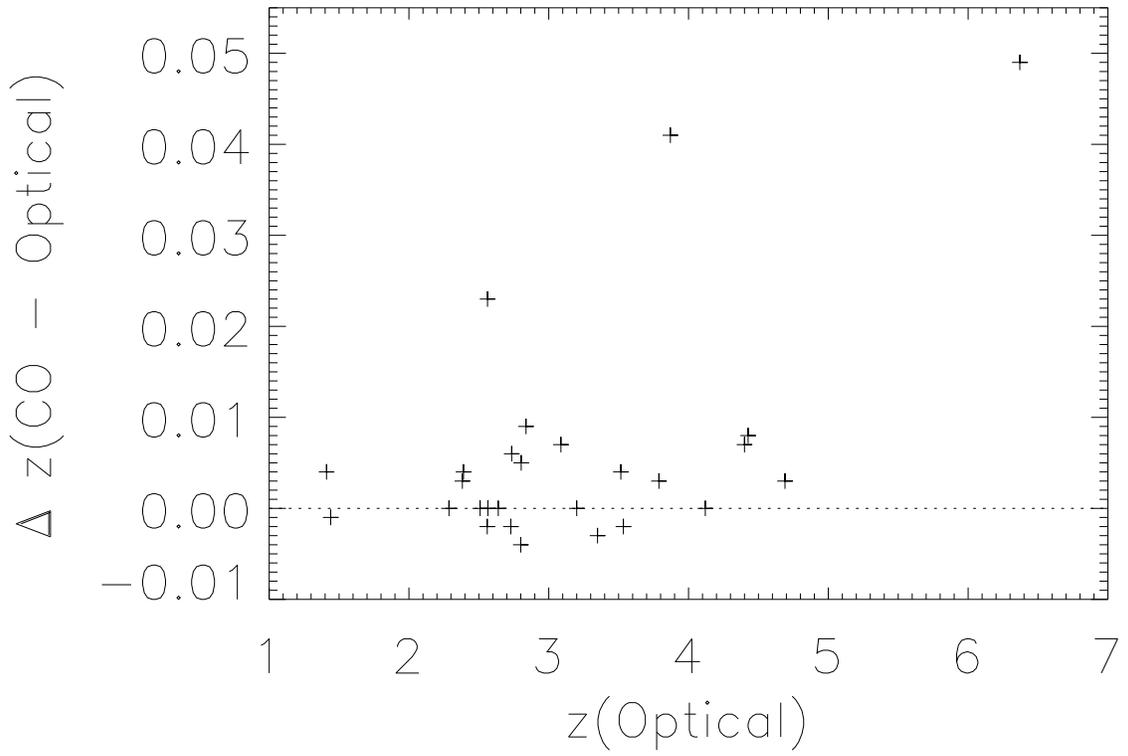}
\caption{Comparison of optical and CO redshifts for high-z CO sources.}
\end{figure}


\clearpage

\begin{thebibliography}{}

\bibitem[Alloin et al.(1992)]{pg1} Alloin, D., Barvainis, R., Gordon, M. A., 
  \& Antonucci, R. R. J.  1992, \aap, 265, 429

\bibitem[Andreani et al.(2000)]{andreani} Andreani, P., Cimatti, A., Loinard, L., 
  \& R\"{o}ttgering, H.  2000, \aap, 354, L1

\bibitem[Bade et al.(1997)]{rxjdisc} Bade, N., Siebert, J., Lopez, S., Voges, W., 
  \& Reimers, D.  1997, \aap, 317, L13

\bibitem[Baker et al.(2004)]{bakerCO} Baker, A. J., Tacconi, L. J., Genzel, R., 
  Lehnert, M. D., \& Lutz, D.  2004, \apj, in press

\bibitem[Barvainis et al.(1994)]{clover} Barvainis, R., Tacconi, L., Antonucci, R., 
  Alloin, D., \& Coleman, P.  1994, \nat, 371, 586

\bibitem[Barvainis et al.(1998)]{bar4} Barvainis, R., Alloin, D., Guilloteau, S., 
  \& Antonucci, R.  1998, \apj, 492, L13

\bibitem[Barvainis et al.(2002a)]{barCOsearch} Barvainis, R., 
  Alloin, D., \& Bremer, M.  2002, \aap, 385, 399

\bibitem[Barvainis \& Ivison(2002b)]{rxjsubmm} Barvainis, R., \& Ivison, R. 
  2002, \apj, 571, 712

\bibitem[Bertoldi et al.(2003a)]{berdust} Bertoldi, F., et al. 2003, \aap, 406, L55

\bibitem[Bertoldi et al.(2003b)]{berCO} Bertoldi, F., et al. 2003, \aap, 409, L47

\bibitem[Blain, Barnard, \& Chapman(2003)]{smmsed} Blain, A. W., Barnard, V. E., 
  \& Chapman, S. C.  2003, \mnras, 338, 733

\bibitem[Burud et al.(1998)]{burud} Burud, I., et al.  1998, \apj, 501, L5

\bibitem[Carilli et al.(2001)]{car2001} Carilli, C. L., et al.  2001, \apj, 555, 625

\bibitem[Carilli et al.(2003)]{psslens} Carilli, C. L., Lewis, G. F., 
  Djorgovski, S. G., Mahabal, A., Cox, P., Bertoldi, F., \&
  Omont, A.  2003, Science, 300, 773

\bibitem[Chapman et al.(2003)]{smmmedian} Chapman, S. C., Blain, A. W., 
  Ivison, R. J., \& Smail, I. R.  2003, \nat, 422, 695

\bibitem[Chavushyan et al.(1995)]{jrussjour} Chavushyan, V. O., Stepanian, J. A.,
  Balayan, S. K., \& Vlasyuk, V. V.  1995, AstL, 21, 804

\bibitem[Condon et al.(1998)]{NVSS} Condon, J. J., Cotton, W. D., Greisen, E. W.,
  Yin, Q. F., Perley, R. A., Taylor, G. B., \& Broderick, J. J.  1998, \aj, 115, 1693

\bibitem[Cowie et al.(2002)]{cowie02} Cowie, L. L., Barger, A. J., \& Kneib, J.--P.  
  2002, \aj, 123, 2197

\bibitem[Cox et al.(2002)]{cox02} Cox, P., et al.  2002, \aap, 387, 40

\bibitem[De Breuck et al.(2003a)]{deBa} De Breuck, C., et al.  2003, \aap, 401, 911

\bibitem[De Breuck et al.(2003b)]{deBb} De Breuck, C., Neri, R., \& Omont, A.
  2003, New Astron. Rev., 47, 285

\bibitem[Dey et al.(1999)]{dey99} Dey, A., Graham, J. R., Ivison, R. J., 
  Smail, I., Wright, G. S., \& Liu, M. C.  1999, \apj, 519, 610

\bibitem[Downes, Solomon, \& Radford(1995)]{dsr} Downes, D., Solomon, P. M., 
  \& Radford, S. J. E.  1995, \apj, 453, L65

\bibitem[Downes \& Solomon(1998)]{dulirg} Downes, D., \& Solomon, P. M.  1998,
  \apj, 507, 615

\bibitem[Downes et al.(1999)]{downes99} Downes, D., Neri, R., Wiklind, T., 
  Wilner, D. J., \& Shaver, P. A.  1999, \apj, 513, L1

\bibitem[Evans et al.(2001)]{pg2} Evans, A. S., Frayer, D. T., Surace, J. A., 
  \& Sanders, D. B.  2001, \aj, 121, 1893 

\bibitem[Ferrarese \& Merritt(2000)]{fm2000} Ferrarese, L., \& Merritt, D. 
  \apj, 539, L9

\bibitem[Foltz et al.(1989)]{lbqsorig} Foltz, C. B., Chaffee, F. H., 
  Hewett, P. C., Weymann, R. J., Anderson, S. F., \& MacAlpine, G. M.
  1989, \aj, 98, 1959

\bibitem[Forster et al.(2001)]{lbqsnal} Forster, K., Green, P. J., 
  Aldcroft, T. L., Vestergaard, M., Foltz, C. B., \& Hewett, P. C.  2001,
  \apjs, 134, 35

\bibitem[Frayer et al.(1998)]{daveCO2} Frayer, D. T., et al.  1998, \apj, 506, L7

\bibitem[Frayer et al.(1999)]{daveCO1} Frayer, D. T., et al.  1999, \apj, 514, L13

\bibitem[Gebhardt et al.(2000)]{gms2000} Gebhardt, K., et al.  2000, \apj, 539, L13

\bibitem[Guilloteau et al.(1997)]{gui97} Guilloteau, S., Omont, A., McMahon, R. G., 
  Cox, P., \& Petitjean, P.  1997, \aap, 328, L1

\bibitem[Guilloteau et al.(1999)]{gui99} Guilloteau, S., Omont, A., Cox, P., 
  McMahon, R. G., \& Petitjean, P.  1999, \aap, 349, 363

\bibitem[Hewett, Foltz, \& Chaffee(1995)]{lbqsz} Hewett, P. C., Foltz, C. B., 
  \& Chaffee, F. H.  1995, \aj, 109, 1498

\bibitem[Holland et al.(1999)]{scuba} Holland, W. S., et al.  1999, \mnras, 303, 659

\bibitem[Hooper et al.(1995)]{lbqsradio} Hooper, E. J., Impey, C. D., 
  Foltz, C. B., and Hewett, P. C.  1995, \apj, 445, 62

\bibitem[Irwin et al.(1998)]{apmoptical} Irwin, M. J., Ibata, R. A., Lewis, G. F., 
  \& Totten, E. J.  1998, \apj, 505, 529

\bibitem[Isaak et al.(2002)]{isaak2002} Isaak, K. G., Priddey, R. S., 
  McMahon, R. G., Omont, A., Peroux, C., Sharp, R. G., \& Withington, S. 
  2002, \mnras, 329, 149

\bibitem[Isaak, Chandler, \& Carilli(2004)]{brhcn}  Isaak, K. G., Chandler, C. J.,
  \& Carilli, C. L.  2004, \mnras, 348, 1035

\bibitem[Ivison et al.(1998)]{ivison98} Ivison, R. J., et al.  1998, \mnras, 
  298, 583

\bibitem[Ivison et al.(2002)]{ivison02} Ivison, R. J., et al.  2002, \mnras, 337, 1

\bibitem[Kneib et al.(1993)]{lenstool} Kneib, J.--P., Mellier, Y., Fort, B., \& 
  Mathez, G.  1993, \aap, 273, 367

\bibitem[Kneib et al.(2000)]{kneib} Kneib, J.--P., Cohen, J. G., 
  \& Hjorth, J.  2000, \apj, 544, L35

\bibitem[Knudsen et al.(2003)]{opticalsmm} Knudsen, K. K., van der Werf, P. P., 
  \& Jaffe, W.  2003, \aap, 411,343

\bibitem[Korista et al.(1993)]{bal93} Korista, K. T., Voit, G. M., Morris,
  S. L., \& Weymann, R. J.  1993, \apjs, 88, 357

\bibitem[Lewis et al.(1998)]{lewis98} Lewis, G. F., Chapman, S. C., Ibata, R. A., 
  Irwin, M. J., \& Totten, E. J.  1998, \apj, 505, L1

\bibitem[McMahon et al.(1999)]{mcmahon99} McMahon, R. G., Priddey, R. S., 
  Omont, A., Snellen, I., \& Withington, S.  1999, \mnras, 309, L1

\bibitem[Neri et al.(2003)]{neri} Neri, R., et al.  2003, \apj, 597, L113

\bibitem[Omont et al.(1996)]{omontCO} Omont, A., Petitjean, P., Guilloteau, S., 
  McMahon, R. G., Solomon, P. M., \& Pecontal, E.  1996, \nat, 382, 428

\bibitem[Omont et al.(2001)]{omont01} Omont, A., Cox, P., Bertoldi, F.,
  McMahon, R. G., Carilli, C., \& Isaak, K. G.  2001, \aap, 374, 371

\bibitem[Omont et al.(2003)]{omont1mm} Omont, A., et al.  2003, \aap, 398, 857

\bibitem[Papadopoulos et al.(2000)]{pap00} Papadopoulos, P., R\"{o}ttgering, H., 
  van der Werf, P. P., Guilloteau, S., Omont, A., van Breugel, W. J. M., \& 
  Tilanus, R.  2000, \apj, 528, 626

\bibitem[Planesas et al.(1999)]{plan} Planesas, P., Martin-Pintado, J., 
  Neri, R., \& Colina, L.  1999, Science, 286, 2493

\bibitem[Priddey et al.(2003a)]{lbqssubmm} Priddey, R. S., Isaak, K. G., 
  McMahon, R. G., \& Omont, A.  2003a, \mnras, 339, 1183 

\bibitem[Priddey et al.(2003b)]{priddey03} Priddey, R. S., Isaak, K. G., 
  McMahon, R. G., Robson, E. I., \& Pearson, C. P.  2003b, \mnras, 344, 74

\bibitem[Richards et al.(1997)]{lbqsmag} Richards, G. T., Yanny, B., Annis, J., 
  Newberg, H. J. M., McKay, T. A., York, D. G., \& Fan, X.  1997, \pasp, 109, 39

\bibitem[Robson et al.(2004)]{sdssdust} Robson, E. I., et al.  2004,
  \mnras, submitted

\bibitem[Rowan--Robinson et al.(1993)]{rr93} Rowan--Robinson, M., et al.  
  1993, \mnras, 261, 513

\bibitem[Sault, Teuben, \& Wright(1995)]{miriadref} Sault, R. J., Teuben,
  P. J., \& Wright, M. C. H. 1995, in ASP Conf. Ser. 77, Astronomical Data
  Analysis Software and Systems IV, ed.\ R. A. Shaw, H. E. Payne, \& J. J. E. 
  Hayes (San Francisco: ASP), 433

\bibitem[Scoville et al.(1993)]{mmaref} Scoville, N. Z., Carlstrom, J. E.,
  Chandler, C. J., Phillips, J. A., Scott, S. L., Tilanus, R. P. J.,
  \& Wang, Z. 1993, \pasp, 105, 1482

\bibitem[Scoville et al.(1997)]{nickCO} Scoville, N. Z., Yun, M. S., Windhorst,
  R. A., Keel, W. C., \& Armus, L.  1997, \apj, 485, L21

\bibitem[Scoville et al.(2003)]{pg3} Scoville, N. Z., Frayer, D. T., 
  Schinnerer, E., \& Christopher, M.  2003, \apj, 585, L105

\bibitem[Skrutskie et al.(1997)]{2mass} Skrutskie, M. F., et al.  1997, Proc.
  Workshop ``The Impact of Large Scale Near-IR Sky Surveys'', ed. Garz\'{o}n, F. 
  et al. (Dordrecht: Kluwer), 25

\bibitem[Smail et al.(1997)]{smail97} Smail, I., Ivison, R. J., \& 
  Blain, A. W.  1997, \apj, 490, L5

\bibitem[Smail et al.(1998)]{smail98} Smail, I., Ivison, R. J., 
  Blain, A. W., \& Kneib, J.--P.  1998, \apj, 507, L21

\bibitem[Smail et al.(2002)]{smail02} Smail, I., Ivison, R. J., 
  Blain, A. W., \& Kneib, J.--P.  2002, \mnras, 331, 495

\bibitem[Smail et al.(2003)]{smail03} Smail, I., Ivison, R. J., Gilbank, D. G., 
  Dunlop, J. S., Keel, W. C., Motohara, K., \& Stevens, J. A.  2003, \apj, 
  583, 551

\bibitem[Solomon, Downes, \& Radford(1992a)]{sdr} Solomon, P. M., Downes, D., 
  \& Radford, S. J. E.  1992a, \nat, 356, 318

\bibitem[Solomon et al.(1992b)]{sdreqpaper} Solomon, P. M., Downes, D., 
  \& Radford, S. J. E.  1992b, \apj, 398, L29

\bibitem[Solomon et al.(2003)]{cloverhcn} Solomon, P., Vanden Bout, P., 
  Carilli, C., \& Guelin, M.  2003, \nat, 426, 636

\bibitem[Stepanian et al.(2001)]{jnewphot} Stepanian, J. A., Green, R. F., Foltz,
  C. B., Chaffee, F., Chavushyan, V. H., Lipovetsky, V. A., \& Erastova, L. K.  2001,
  \aj, 122, 3361

\bibitem[Tytler \& Fan(1992)]{tytlerfan} Tytler, D., \& Fan, X.  1992, \apjs,
  79, 1

\bibitem[Walter et al.(2003)]{fabian_sdss} Walter, F., et al. 2003, \nat, 424, 406 

\bibitem[White et al.(1997)]{first} White, R. L., Becker, R. H., Helfand, D. J., 
  \& Gregg. M. D.  1997, \apj, 475, 479

\bibitem[Yun \& Carilli(2002)]{MinSED} Yun, M. S., \& Carilli, C. L. 2002,
  \apj, 568, 88

\bibitem[Yun et al.(2004)]{minnew} Yun, M. S., Reddy, N. A., Scoville, N. Z., 
  Frayer, D. T., Robson, E. I., \& Tilanus, R. P. J.  2004, \apj, 601, 723

\end{thebibliography}
\end{document}